\documentclass{iaus}
\usepackage{graphicx, natbib}

\title[Carbon Abundances in SMC PNe] %% give here short title %%
{Carbon Abundances in the Small Magellanic Cloud Planetary Nebulae}

\author[Lee et al.]   %% give here short author list %%
{Ting-Hui Lee$^1$, L. Stanghellini$^1$, R. A. Shaw$^1$, B. Balick$^2$ 
 \break \and E. Villaver$^3$}

\affiliation{$^1$National Optical Astronomy Observatory, 950 North
  Cherry Avenue, Tucson, AZ 85719, USA \\[\affilskip]
$^2$Department of Astronomy, University of Washington, Seattle, WA
  98195, USA  \\[\affilskip]
$^3$Space Telescope Science Institute, 3700 San Martin Drive,
  Baltimore, MD 21218, USA}

\pubyear{2006}
\volume{234}  %% insert here IAU Symposium No.

\begin{document}

\maketitle

\begin{abstract}

As an ongoing study of Magellanic Cloud PNe we have obtained UV
spectra of 9 PNe in the SMC to measure their carbon abundances.  The
spectra have been acquired with ACS HRC/PR200L and SBC/PR130L.  The
ACS prisms give a reasonable resolution in the range of 1200 -- 2500
\AA~to detect the C~{\footnotesize IV}, C~{\footnotesize III]}, and
C~{\footnotesize II]} nebular emission, essential for chemical studies
of the PNe.  The carbon abundances of SMC PNe, together with those of
the LMC previously determined with STIS spectroscopy, will allow a
comparative study of nebular enrichment and provide the basis for
comparison with stellar evolution models at various metallicity.

\end{abstract}

Planetary Nebulae (PNe) in the Magellanic Clouds, with known distances
and relatively low interstellar extinction, are free of two major
biases that hinder Galactic PNe studies, thus they offer a unique
opportunity to study the evolution of low- and intermediate-mass
stars.

Recently, the carbon abundances of 24 LMC PNe have been determined
with the UV spectra acquired with the HST/STIS.  The average carbon
abundance in round and elliptical PNe is larger than that of the
bipolar PNe, confirming that bipolarity in LMC PNe is tightly
correlated with high-mass progenitors \citep{stanghellini05}.

\begin{figure}
\includegraphics[width=135mm]{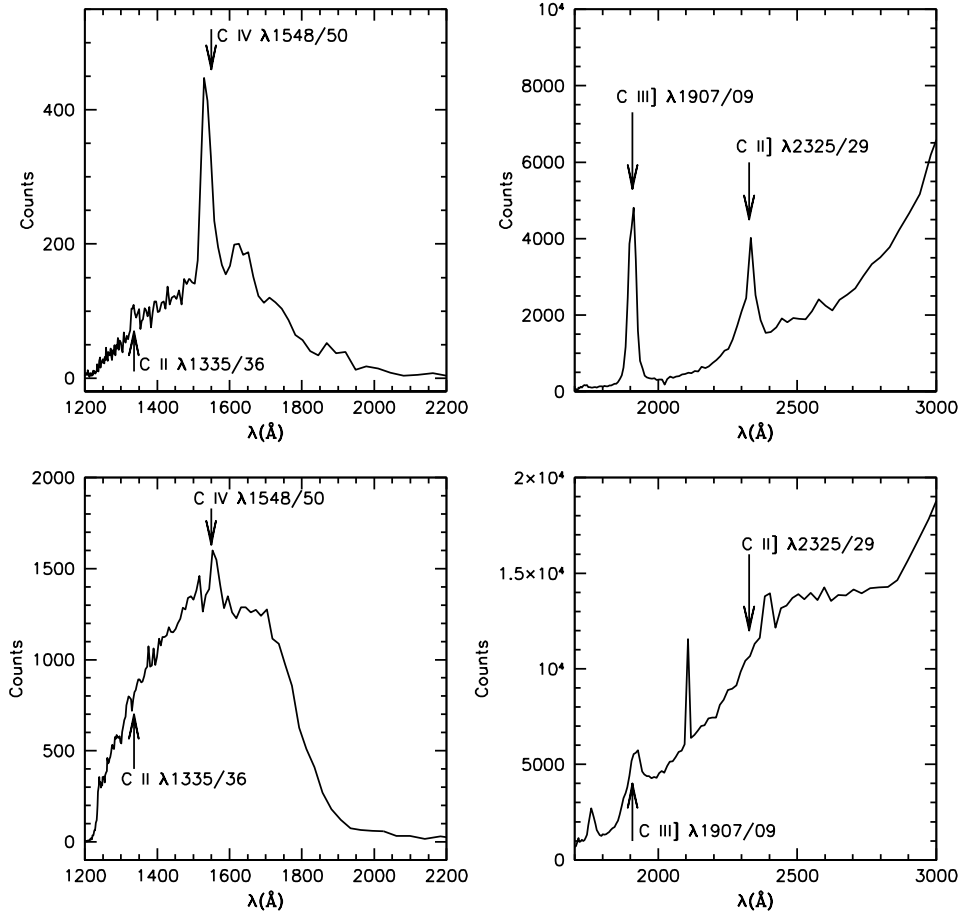}
\caption{Top: Spectrum of SMP 6 obtained with SBC (left) and HRC
  (right).  Bottom: Spectrum of SMP 8 obtained with SBC (left) and HRC
  (right).  The carbon lines are indicated with arrows.}
\label{fg:smcpn}
\end{figure}

Planetary Nebulae in the Magellanic Clouds show similar morphologies
as the Galactic PNe: round, elliptical, and bipolar.  If we compare
PNe in the LMC and the SMC, we note that the fraction of aspherical
PNe in the LMC is higher than in the SMC.  This suggests that the
low-metallicity environment of the SMC discourages shaping bipolar PNe
\citep{stanghellini03}.  To study the chemical enrichment in such
environments, we have acquired UV spectra of SMC PNe with the HST/ACS.
The ACS HRC/PR200L and SBC/PR130L prisms give a reasonable resolution
in the range of 1200 -- 2500 \AA~to detect the C~{\footnotesize IV},
C~{\footnotesize III}], and C~{\footnotesize II}] nebular emission.
The UV spectra of two PNe are shown in Figure \ref{fg:smcpn}.

In addition to these two PNe, we have also obtained spectra for SMP 13,
15, 16, 18, 20, 26, and 28, and SMP 24 and 26 will be observed in
June-August.  We are currently working on the calibration and
extraction of the spectra in order to derive the carbon abundances of
these PNe.  The central star properties and morphology of these PNe
have been previously determined \citep{villaver04, stanghellini03}.
We will derive the abundance-to-mass relationship, and correlate it
with their morphology.  This will allow us to probe the cosmic
recycling, test the PN luminosity function, and study the stellar
population and evolution in a very low-metallicity environment.

\clearpage
\begin{acknowledgments}

Support for program GO-10259 is provided by NASA through a grant from
the Space Telescope Science Institute, which is operated by the
Associate of Universities for Research in Astronomy, Inc., under NASA
contract NAS 5-26555.

\end{acknowledgments}

\end{document}